\documentclass[conference]{IEEEtran}
\usepackage{cite}

\ifCLASSINFOpdf
\else
   \usepackage[dvips]{graphicx}
   \graphicspath{{../}}
   \DeclareGraphicsExtensions{.eps}
\fi
\begin{document}
%
\title{ Utility-maximization Resource Allocation for Device-to-Device Communication Underlaying Cellular Networks }

\author{\IEEEauthorblockN{Jialiang Zhang, Gang Wu, Wenhui Xiong, Zhi Chen and Shaoqian Li}
\IEEEauthorblockA{National Key Laboratory of Science and Technology on Communications,
\\University of Electronic Science and Technology of China,Chengdu,China}
}
\setlength{\arraycolsep}{0.0em}

%


\maketitle

\begin{abstract}
Device-to-device(D2D) underlaying communication brings great benefits to the cellular networks from the improvement of coverage and spectral efficiency at the expense of complicated transceiver design. With frequency spectrum sharing mode, the D2D user generates interference to the existing cellular networks either in downlink or uplink. Thus the resource allocation for D2D pairs should be designed properly in order to reduce possible interference, in particular for uplink. In this paper, we introduce a novel bandwidth allocation scheme to maximize the utilities of both D2D users and cellular users. Since the allocation problem is strongly NP-hard, we apply a relaxation to the association indicators. We propose a low-complexity distributed algorithm and prove the convergence in a static environment. The numerical result shows that the proposed scheme can significant improve the performance in terms of utilities.The performance of D2D communications depends on D2D user locations, the number of D2D users and QoS(Quality of Service) parameters.
\end{abstract}


%
\IEEEpeerreviewmaketitle

\section{Introduction}
To meet increasing demand of wireless data service and shortage of frequency spectrum, Device-to-device(D2D) underlaying cellular network have been proposed. D2D communications allow proximity users that in a cellular network can communicate directly. Because of the shorter transmitting distance, D2D communications can use lower transmitting power to reduce the intra-cell interference and the power consumption of mobile devices. Otherwise, D2D communications can unload parts of the traffic undertaken by the BS. D2D communications underlaying cellular networks is a competitive technique of increasing the access probability in the area with high density like concerts,offices and campuses\cite{janis2009device}.

In \cite{doppler2010mode}, the authors concluded two basic mode of D2D communication: orthogonal mode and reuse mode. The orthogonal mode is  discussed in \cite{fitzek2009cellular}. The resources are allocated to the cellular users and D2D users with no overlapping. This sharing method is easy to implement because there is no interference between cellular users and D2D users.  But the orthogonal mode couldn't bring much improvement in a fully loaded net.So the reusing mode is discussed wider in \cite{yu2011resource,yu2009power,yu2009performance}. The D2D communications share the same resources to achieve high spectrum efficiency. However, the reusing may cause serious interference problem in the network. So one of the key issue in D2D communication underlaying cellular networks is the resource allocation problem which affect both the network capacity and quality of service.

In this paper, we propose the resource allocation problem in D2D communications underlaying cellular networks that maximize the sum of D2D users' utilities with the constraints of the decreased utilities of cellular users caused by the interference from D2D users. This proposed problem is a binary optimization problem which is a strongly NP-hard problem and the corresponding solution may be not practically. So we relax the binary association indicator into continuous variables, and the problem turns out into a convex optimization problem. We solve the convex problem via primal-dual decomposition which can converges to the near-optimal solution. This provide a efficient and low-overhead scheme for implementation.

The rest of paper is organized as follows: In Section II, we address the system model of D2D communications underlaying cellular networks in reuse mode and the definition of utility function. In section III, We proposed the optimization problem and relax it into a convex optimization problem.In Section IV, we develop the distributed resource allocation algorithm via primal-dual decomposition. In section V, we present several numerical results. Finally,we give the conclusion in Section VI.

\section{System Model}
In this section, we describe the model of D2D communication and utility function used in this paper.
\subsection{Model of D2D Communication Reusing the Uplink Resources}
We consider a single cell scenario, in which multiple UEs are uniformly distributed within the cell. In general, uplink spectrum in frequency-division duplex (FDD) is often under-utilized than that of downlink. D2D communication with downlink frequency resue mode is less feasible\cite{zulhasnine2010efficient}. In this paper, we consider D2D communication using uplink resource, i.e., the D2D user is capable to detect uplink signal. Taking 3GPP LTE cellular networks as an example, a D2D UE is also equipped with receiver for single carrier frequency-division multiple-access (SC-FDMA) signal.

There are ${N_{cellular}}$ active cellular UEs and ${N_{D2D}}$ potential D2D pairs which includes one transmitter and one receiver. We assume the cellular network is fully loaded similar to \cite{janis2009interference}. The spectrum is divided into N orthogonal channels and  the bandwidth of ${j}$th channel is ${W\left(j\right)}$  and there is no spare spectrum. We use ${S_C}$ and ${S_{D2D}}$ to indicate the index sets of cellular UEs and D2D pairs.

\begin{figure}[!t]
\centering
\includegraphics[width=3in]{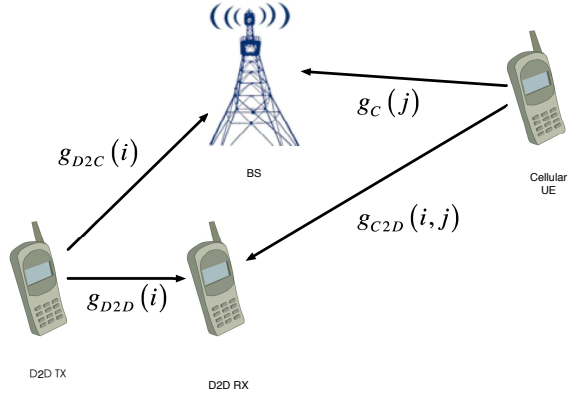}
\caption{System model of D2D communications sharing uplink resources of Single Cell.Only one D2D pair and one cellular user is shown for simplicity. }
\label{fig_model}
\end{figure}

As shown in Fig.\ref{fig_model}, ${g_c\left( j \right)}$ is the channel gain between ${j}$th cellular UE and BS, ${g_{D2D}\left( i \right)}$ is the channel gain between ${i}$th D2D transmitter and receiver. ${g_{D2C}\left( i \right)}$ is the channel gain between ${i}$th D2D transmitter and BS,and ${g_{C2D}\left( i,j \right)}$ is the channel gain between ${i}$th D2D transmitter and ${j}$th cellular UE.

We adopt a pathloss dominated additive white Gaussian noise channel with log-normal shadowing. Thus, all the channel gain can be expressed as
\begin{equation}
g = K \cdot {L^{ - \alpha }} \cdot \xi
\end{equation}
where $K$ is a constant determined by the system parameter, $\alpha$ is the pathloss exponent, $L$ is the distance between devices and $\xi$ is the shadowing component which follows the log-normal distribution.

We apply Shannon capacity to both Cellular UEs and D2D pairs as below.
\begin{eqnarray}
\label{eq11}
&&{r_c}\left( j \right) = W\left(j\right)\log \left( {1 + \frac{{{P_C}{g_C}\left( j \right)}}{{{P_N}}}} \right)
\end{eqnarray}

\begin{eqnarray}
&&{r_c}'\left( j \right) = \nonumber\\
&&W\left(j\right)\log \left( {1 + \frac{{{P_C} {g_C}\left( j \right)}}{{\sum\limits_{i \in {S_{D2D}}} {x\left( {i,j} \right){P_{D2D}}{g_{D2C}}\left( i \right)}  + {P_N}}}} \right)\label{eq22}
\end{eqnarray}

\begin{equation}
{r_{D2D}}\left( {i,j} \right) = W\left(j\right)\log \left( {1 + \frac{{{P_{D2D}}{g_{D2D}}\left( i \right)}}{{{P_C}{g_{C2D}}\left( {i,j} \right) + {P_N}}}} \right)
\end{equation}

where ${r_c}\left( j \right)$ is the spectrum efficiency of the ${j}$th cellular UEs without interference from D2D UEs, ${r_c}'\left( j \right)$is the spectrum efficiency of the ${j}$th cellular UEs with interference from D2D UEs, ${r_{D2D}}\left( {i,j} \right)$ is the spectrum efficiency of ${i}$th D2D pair when it shares the resource of ${j}$th cellular UE. ${P_{D2D}}$,${P_{cellular}}$,${P_{N}}$ is the power density of D2D UEs, cellular UEs and noise.

$x\left( {i,j} \right)$ is the association indicator defined as below.
\begin{equation}
x\left( {i,j} \right) = \left\{ {\begin{array}{lcl}
{1,\mbox{D2D user }i\mbox{ associated with cellular user } j}\\
{0,\mbox{others}}
\end{array}} \right.
\end {equation}

\subsection{Utility Function}
Utility function was first introduced in micro economics and used in the data network for cross layer optimization and balancing the efficiency and fairness\cite{song2005cross}.
The utility function should follow the following assumption:

C1:the utility functions are increasing, strictly concave, and twice continuously differentiable.

C2:The curvatures of utility functions are bounded away from zero.

The utility function for best effort traffic used in this paper is proposed in \cite{jiang2005max}:
\begin{equation}
U\left( r \right) = 0.16 + 0.8\ln \left( {r - 0.3} \right)
\end{equation}

In\cite{jiang2005max},the Unit of $r$ is $k bps$, but in the past decades,the data rate of wireless data network has been increased several times. We simply change the unit to $M bps$ to capture the user experience nowadays.

\section{Problem Formulation}
In this section we state the optimization problem and its relaxation leads to our resource allocation framework. A distributed algorithm will be discussed in the following section.
\subsection {Utility Maximization Problem}
D2D communications underlaying cellular networks could be used to improve the throughput of existing cellular network. The incurred interference to the cellular UEs caused by the D2D UEs should be below a threshold. In this case we investigate a optimization problem, which is to maximize the sum utility of D2D UEs and the decreased utilities of each cellular UE should not exceed a given margin $\beta$. The utility optimization problem can be formulated as follows.
\begin{equation}
\begin{array}{ll}
\mathop {\arg\max }\limits_x {\rm{ }}&\sum\limits_{i \in {S_{D2D}}} {{U_{D2D}}\left( {\sum\limits_{j \in {S_C}} {x\left( {i,j} \right){r_{D2D}}\left( {i,j} \right)} } \right)} \\
\mbox{s.t.}&{U_c}\left( {{r_c}\left( j \right)} \right) - {U_c}\left( {{r_c}'\left( j \right)} \right) \le \beta,\forall j \in {S_c} \nonumber\\
&\sum\limits_{i \in {S_{D2D}}} {x\left( {i,j} \right) \le 1{\rm{  }},\forall j \in {S_c}} \nonumber\\
&x\left( {i,j} \right){\rm{ = }}\left\{ {0,1} \right\}{\rm{ }}\forall i \in {S_{D2D}},j \in {S_C}{\rm{ }}\nonumber
\label{problem1}
\end{array}
\end{equation}
The problem formulated in (\ref{problem1}) is a strongly  NP-hard combinational optimization  problem. The computing complexity is $\mathcal{O}\left({N_{D2D}}^{N_{Cellular}}\right)$, where ${N_{D2D}}$ and ${N_{Cellular}}$ is the number of active cellular UEs and D2D pairs. The complexity increases rapidly when the number of cellular UEs and D2D UEs grows. Therefore, we relax the binary variable $x\left(i,j\right)$ and make the problem (\ref{problem1}) a convex problem below.
\subsection{Relaxation of the association indicator }
We relax the association indicator $x\left(i,j\right)$ into continuous variables between 0 and 1 which means the D2D pair could reuse part of the resource of a cellular UE. The indicator $x\left(i,j\right)$ is now a fraction and the relaxation is formulated below.
\begin{equation}
\begin{array}{ll}
\mathop {\arg\max }\limits_x {\rm{ }}&\sum\limits_{i \in {S_{D2D}}} {{U_{D2D}}\left( {\sum\limits_{j \in {S_C}} {x\left( {i,j} \right){r_{D2D}}\left( {i,j} \right)} } \right)} \\
\mbox{s.t.}&{U_c}\left( {{r_c}\left( j \right)} \right) - {U_c}\left( {{r_c}'\left( j \right)} \right) \le \beta,\forall j \in {S_c} \nonumber\\
&\sum\limits_{i \in {S_{D2D}}} {x\left( {i,j} \right) \le 1{\rm{  }},\forall j \in {S_c}} \nonumber\\
&{\rm{0}} \le x\left( {i,j} \right) \le {\rm{1,}}\forall i \in {S_{D2D}},j \in {S_C}
\label{problem2}
\end{array}
\end{equation}
The above relaxation (\ref{problem2}) is convex\cite{boyd2004convex}.To solving this problem using a centralized way need the global channel gains which increase exponentially when the number of devices grows up. So we introduce a distributed algorithm to reduce the complexity and the cost of signaling interaction.
\section{Synchronous Distributed Algorithm }
In this section, we first derive a transform of problem (\ref{problem2}) and its Lagrangian dual which could be decomposed into ${N_{D2D}}$ separable subproblem. Then we investigate a distributed Algorithm to solve the problem using gradient method. Finally, we discuss the dynamic step size and its convergence.
\subsection{Primal-Dual Decomposition}
Primal-dual decomposition is a general approach to solving a problem by breaking it up into smaller ones and solving each of the smaller ones separately in a parallel and it was widely used in the data network flow control \cite{boyd2007notes} and load balancing in heterogeneous cellular network\cite{6503766}. We first transform the the first constraint of problem (\ref{problem2}) into the standard form, as below.
\begin{eqnarray}
&\mathop {\arg\max }\limits_x {\rm{ }}&\sum\limits_{i \in {S_{D2D}}} {{U_{D2D}}\left( {\sum\limits_{j \in {S_C}} {x\left( {i,j} \right){r_{D2D}}\left( {i,j} \right)} } \right)} \\
&\mbox{s.t.}&\sum\limits_{i \in D2D} {{P_{D2D}}\left( i \right){g_{D2D}}\left( {i,j} \right)} x\left( {i,j} \right) \le   \beta \left( j \right),\forall j \in {S_c}\nonumber\\
&&\sum\limits_{i \in {S_{D2D}}} {x\left( {i,j} \right) \le 1{\rm{  }},\forall j \in {S_c}} \label{c2}\\
&&{\rm{0}} \le x\left( {i,j} \right) \le {\rm{1,}}\forall i \in {S_{D2D}},j \in {S_C}\nonumber
\end{eqnarray}
where
\begin{equation}
\beta \left( j \right)=\frac{{{P_c}{g_c}\left( j \right)}}{{{2^{\frac{{U_c^{ - 1}\left( {{U_c}\left( {{r_c}\left( j \right)} \right) - \beta } \right)}}{{W(j)}}}} - 1}}
\end{equation}
The first two constraints of (\ref{c2}) are coupling constraints which make us to turn to primal-dual decomposition by adding two Lagrange multiplier $f_{1}\left(j\right)$ and $f_{2}\left(j\right)$.

Define the Lagrangian

\begin{eqnarray}
&&L\left( {x,p} \right) = \sum\limits_{i \in {S_{D2D}}} {{U_{D2D}}\left( {\sum\limits_{j \in {S_c}} {x\left( {i,j} \right){r_{D2D}}\left( {i,j} \right)} } \right)} \nonumber \\
&&- \sum\limits_{j \in {S_c}} {{p_1}\left( j \right)\left( {\sum\limits_{i \in D2D} {x\left( {i,j} \right){g_{D2C}}\left( i \right)}  - \beta \left( j \right)} \right)} \nonumber \\
&&- \sum\limits_{j \in Sc} {{p_2}\left( j \right)\left( {\sum\limits_{i \in D2D} {x\left( {i,j} \right) - 1} } \right)} \\
&& = \sum\limits_{i \in {S_{D2D}}}  {U_{D2D}}\left( {\sum\limits_{j \in {S_c}} {x\left( {i,j} \right){r_c}\left( {i,j} \right)} } \right) \label{L1}\nonumber\\
&&- \sum\limits_{i \in {S_{D2D}}}\left({g_{D2C}}\left( i \right)\sum\limits_{j \in {S_c}} {{p_1}\left( j \right)x\left( {i,j} \right)- \sum\limits_{j \in {S_c}} {{p_1}\left( j \right)x\left( {i,j} \right)} } \right)\label{L2}\nonumber\\
&&+ \sum\limits_{j \in {S_c}} {{p_1}\left( j \right)\beta \left( j \right)}  + \sum\limits_{j \in {S_c}} {{p_2}\left( j \right)\label{L3}}
\end{eqnarray}
Notice the first two componenents (\ref{L3}) are separable in $x\left(i,j\right)$  so we can derive (\ref{eqa}).
\newcounter{mytempeqncnt}
\begin{figure*}[!t]
\normalsize
\begin {eqnarray}
\mathop {\max }\limits_x \sum\limits_{i \in {S_{D2D}}} {\left( {U\left( {\sum\limits_{j \in {S_C}} {x\left( {i,j} \right){r_{D2D}}\left( {i,j} \right)} } \right) - {g_{D2C}}\left( i \right)\sum\limits_{j \in {S_c}} {{p_1}\left( j \right)x\left( {i,j} \right) - \sum\limits_{j \in {S_c}} {{p_1}\left( j \right)x\left( {i,j} \right)} } } \right)} \nonumber\\
=\sum\limits_{i \in {S_{D2D}}} {\mathop {\max }\limits_x \left( {U\left( {\sum\limits_{j \in {S_C}} {x\left( {i,j} \right){r_{D2D}}\left( {i,j} \right)} } \right) - {g_{D2C}}\left( i \right)\sum\limits_{j \in {S_c}} {{p_1}\left( j \right)x\left( {i,j} \right) - \sum\limits_{j \in {S_c}} {{p_1}\left( j \right)x\left( {i,j} \right)} } } \right)} \label{eqa}
\end{eqnarray}
\vspace*{4pt}
\end{figure*}

The objective function of the dual problem is

\begin{equation}
D\left( p \right) = \mathop {\max }\limits_x L\left( {x,p} \right) = \sum\limits_{i \in {S_{D2D}}} {{f_i}\left( p \right)}  + g\left( p \right),
\end{equation}
where ${f_i}\left( p \right)$ and $g\left( p \right)$ is given in (17) and (18)

\begin{figure*}[!t]
\normalsize
\begin{eqnarray}
&&{f_i}\left( p \right) = \left\{ {\begin{array}{*{20}{c}}
{\mathop {\max }\limits_{{x_i}} {\rm{ }}{U_{D2D}}\left( {\sum\limits_{j \in Sc} {x\left( {i,j} \right){r_{D2D}}\left( {i,j} \right)} } \right) - {g_{D2C}}\left( i \right)\sum\limits_{j \in {S_C}} {x\left( {i,j} \right){p_1}\left( j \right)}  - \sum\limits_{j \in {S_C}} {x\left( {i,j} \right){p_2}\left( j \right)} }\label{subproblem}\\
\begin{array}{l}
{\rm{  }}\\
{\rm{ s.t.      }}0 \le x\left( {i,j} \right) \le 1
\end{array}
\end{array}} \right.\\
&&g\left( p \right) = \sum\limits_{j \in {S_c}} {{p_1}\left( j \right)\beta \left( j \right)}  + \sum\limits_{j \in {S_c}} {{p_2}\left( j \right)}
\end{eqnarray}
\hrulefill
\vspace*{4pt}
\end{figure*}
and the dual problem is
\begin{equation}
\label{D1}
\mathbf{D:}\mathop {\min }\limits_{p > 0} D\left( p \right)
\end{equation}

The constraints (\ref{c2}) are all linear equalities and inequalities hence the refined Slater condition reduces to feasibility and the strong duality holds \cite{boyd2004convex} which means we can solve primal problem equivalently by solving the dual problem  (\ref{D1}). As the dual optimal $p_1^*$ and $p_2^*$, the primal problem can be solved separately by each D2D pairs without any coordination.

\subsection{Distributed Algorithm}
We solve the dual problem by projected gradient method\cite{bertsekas1989parallel,low1999optimization,6503766}.Even the gradient method is much slower than interior-point methods,but it could combined with the primal-dual decomposition technique to develop a simple distributed algorithm\cite{boyd2007notes}. In our problem, the distributed way brings a advantage that the BS only have to acquire $g_c$,and$g_{D2C}$ which could be measured directly compared to all the channel information in a centralized way.

The Lagrangian  multiplier $p_1\left(j\right)$ and $p_2\left(j\right)$ which could be considered as the prices of reusing the resource of cellular UE $\left(j\right)$ adjusted in the opposite direction to the gradient $\bigtriangledown D\left(p\right)$.
 \begin{equation}
\label {p11}
{p_{1,t + 1}}\left( j \right) = {\left[ {{p_{1,t}}\left( j \right) - {\gamma _1}\frac{{\partial D}}{{\partial {p_1}\left( j \right)}}\left( p \right)} \right]^ + }
\end{equation}
 \begin{equation}
\label {p21}
{p_{2,t + 1}}\left( j \right) = {\left[ {{p_{2,t}}\left( j \right) - {\gamma _2}\frac{{\partial D}}{{\partial {p_2}\left( j \right)}}\left( {{p_2}} \right)} \right]^ + }
\end{equation}
where ${\gamma _1}$,${\gamma _2}$ is positive step size and ${[z]^ + } = \max \left\{ {z,0} \right\}$.

Since the utility $U(x)$ is strictly concave,$D\left(p\right)$ is continuously differentiable with derivatives given by
\begin{eqnarray}
&&\frac{{\partial D}}{{\partial {p_1}\left( j \right)}}\left( {{p_1}} \right) =  - \sum\limits_{i \in {S_{D2D}}} {{g_{D2C}}\left( i \right)x\left( {i,j} \right) + \beta \left( j \right)}\label{diff1} \\
&&\frac{{\partial D}}{{\partial {p_2}\left( j \right)}}\left( {{p_2}} \right) =  - \sum\limits_{i \in {S_{D2D}}} {x\left( {i,j} \right) + 1}\label{diff2}
\end{eqnarray}
Substituting (\ref{diff1}) into (\ref{p11}), (\ref{diff2}) into (\ref{p21}), we obtain the following price adjustment rules for cellular UEs
\begin {eqnarray}
&&{p_{1,t + 1}}\left( j \right) = \nonumber\\
&&{\left[ {{p_{1,t}}\left( j \right) + {\gamma _1}\left( {\sum\limits_{i \in {S_{D2D}}} {{g_{D2C}}\left( i \right)x\left( {i,j} \right) - \beta \left( j \right)} } \right)} \right]^ + }\label{p12}\\
&&{p_{2,t + 1}}\left( j \right) = {\left[ {{p_{2,t}}\left( j \right) + {\gamma _2}\left( {\sum\limits_{i \in {S_{D2D}}} {x\left( {i,j} \right) - 1} } \right)} \right]^ + }\label{p22}
\end {eqnarray}

If we treat ${\sum\limits_{i \in {S_{D2D}}} {x\left( {i,j} \right) } }$ as the demand of resources  and ${\sum\limits_{i \in {S_{D2D}}} {{P_{D2D}}{g_{D2C}}\left( i \right)x\left( {i,j} \right)} }$ as the demand of allowance of interference of cellular UE $j$ and. A constant 1 and $\beta\left(j\right)$ are the service the cellular UE $j$ can provide. Then the $p_1\left(j\right)$,and$f_2\left(j\right)$ are the bridges between demand and supply. Equation(\ref{p12})(\ref{p22}) is consistent with the law of supply and demand: if the demand
${\sum\limits_{i \in {S_{D2D}}} {x\left( {i,j} \right) } }$  exceeds the supply at cellular UE $j$, raise  $p_1\left(j\right)$,and$f_2\left(j\right)$ ;otherwise reduce $p_1\left(j\right)$,and$f_2\left(j\right)$. Notice the adjustment algorithm can be implemented by using only local information of BS.

Then we consider the maximization subproblem (\ref{subproblem}) based on the fact that the resource of uplink can only be allocated continuously in frequency in the practical cellular system. We formulate the problem (\ref{subproblem})into
\begin{eqnarray}
\mathop {\max }\limits_x {\rm{ }}{U_{D2D}}\left( {\sum\limits_{j \in Sc} {\left( {x\left( {i,j} \right)W\left( j \right){r_{D2D}}\left( {i,j} \right)} \right)} } \right)\nonumber\\
{ - {g_{D2C}}\left( i \right)\sum\limits_{j \in {S_C}} {x\left( {i,j} \right){p_1}\left( j \right)}  - \sum\limits_{j \in {S_C}} {x\left( {i,j} \right){p_2}\left( j \right)} }
\end{eqnarray}
which means a D2D pair can only reuse the resource of one cellular UE with a ration between 0 and 1. We can get the maximizer from the Karush–Kuhn–Tucker(KKT) condition. The numerical result shows that the loss can be ignored when the number of users is large enough. Besides the reduction of complexity, we can also benefit from the reduction of the feedback information to BS from the D2D pairs at each iteration from $N_{D2D}\times N_{Cellular}$ to $N_{D2D}$. We summarize the algorithm as follows:

\textbf{1)BS Algorithm:}

i) BS initializes the the prices $p_1\left(j\right)=1$, and $p_2\left(j\right)=1$ measure the channel gains , ${g_c\left( j \right)}$and, ${g_{D2C}\left( i \right)}$, calculates the $\beta\left(j\right)$ and sends the feedback information ${g_{D2C}\left( j \right)}$ to D2D pair $i$ at the beginning.

ii) At each iteration, the BS receives the chosen reusing resources and reusing ratios feedback of each D2D pair. The prices $p_1\left(j\right)$,and$p_2\left(j\right)$ are updated by (\ref{p12}) and (\ref{p22}).

\textbf{2)D2D pairs' Algorithm:}

i) Each D2D pair measures the channel gains ${g_{D2D}\left( i \right)}$ and ${g_{C2D}\left( i,j \right)}$ and receives the channel gains ${g_{D2C}\left( i \right)}$ from BS at the beginning.

ii) At each iteration, each D2D pair receives the prices broadcast by BS and determines which cellular UE's resource it will reuse with and the reusing ratio by the following equation.
\begin{eqnarray}
&&(x^*,j^*)=\arg\mathop {\max }\limits_{x,j} {U_{D2D}}\left( {\sum\limits_{j \in Sc} { {x\left( {i,j} \right)W\left( j \right){r_{D2D}}\left( {i,j} \right)}} } \right)\nonumber\\
&&{ - {g_{D2C}}\left( i \right)\sum\limits_{j \in {S_C}} {x\left( {i,j} \right){p_1}\left( j \right)}  - \sum\limits_{j \in {S_C}} {x\left( {i,j} \right){p_2}\left( j \right)} }
\end{eqnarray}
\subsection{Step Size and Convergence Analysis}
The stepsize of both $p_1\left(j\right)$,and$p_2\left(j\right)$ is updated following the rule below.
\begin{eqnarray}
\gamma (t) = \alpha \left( t \right)\frac{{D\left( {p\left( t \right)} \right)
 - D\left( t \right)}}{{{{\left\| {\partial D\left( {p\left( t \right)} \right)} \right\|}_2}}},0 < \alpha \left( t \right) < 2
\end{eqnarray}
where $D\left(t\right)$ is an estimate of optimal value $D^*$ of problem (\ref{D1}). The procedure for updating $D\left(t\right)$ is as follows
\begin{equation}
D\left( t \right) = \mathop {\min }\limits_{k \in {S_c}} D\left( {p\left( k \right)} \right) - \varepsilon \left( t \right)
\end{equation}
and $\varepsilon \left( t \right)$ is updated by
\begin{equation}
\varepsilon \left( {t + 1} \right) = \left\{ {\begin{array}{*{20}{c}}
{\max \left\{ {\rho \varepsilon \left( t \right),\varepsilon } \right\},D\left( {{p_1}\left( {t + 1} \right)} \right) \ge D\left( {p\left( t \right)} \right)}\\
{\rho \varepsilon \left( t \right),D\left( {p\left( {t + 1} \right)} \right) < D\left( {{p_1}\left( t \right)} \right)}
\end{array}} \right.
\end{equation}
where $\beta <1$ and $\rho >1$.

In this procedure, we want to reach a target level $D\left(t\right)$ that is smaller by $\varepsilon \left( t \right)$ over the best value achieved thus far.Whenever the target level is achieved we increase $\varepsilon \left( t \right)$. If the target level is not attained at a given iteration, $\varepsilon \left( t \right)$ up to a threshold $\varepsilon$. This threshold guarantees that the stepsize is bounded away form zero. As the  derivative of function $D\left(p\right)$ given by  (\ref{diff1})(\ref{diff2}) are all bounded, the subgradient of dual objective function is also bounded:
\begin{equation}
\mathop {\sup }\limits_t \left\{ {\left\| {\partial D\left( {\mu \left( t \right)} \right)} \right\|} \right\} \le c
\end{equation}
where $c$ is some scalar and it satisfies the assumption of Proposition 6.3.6 in \cite{bertsekas2009convex}. By applying the proposition, the dynamic stepsize adjustment procedure leads to converges to a  near optimal solution.
\section{Numerical Result and analysis}
We consider single cell scenario,where cellular UEs is uniformly distributed in the cell. We apply the clustered model where D2D pairs uniformly distributed in a randomly located cluster with given radius.Other simulation parameters are summarized in Table.\ref{table}. Water-filling method of utilities is used to allocate the resource of cellular users.
\begin{table}[!t]
\renewcommand{\arraystretch}{1.3}
\caption{Simulation Parameters}
\label{table}
\centering
\begin{tabular}{|p{1.25in}|p{1.25in}|}
\hline
Parameter & Value\\
\hline
\hline
Cell Radius & 500m\\
\hline
Uplink bandwidth& 5MHz\\
\hline
Noise Power Density&-174dBm/Hz\\
\hline
Pathloss exponent&4\\
\hline
Pathloss constant&24dBm\\
\hline
Shadowing stand deviation of cellular link&8dB \\ \hline
Shadowing stand deviation of D2D link&6dB \\
\hline
D2D pair radius&10,20,40,80,160 m\\
\hline
Number of cellular users& 50\\
\hline
Number of D2D pairs&20,25,40,50\\
\hline
Qos constraint $\beta$ &0.5\\
\hline
\end{tabular}
\end{table}

\begin{figure}[!t]
\centering
\includegraphics[width=2.5in]{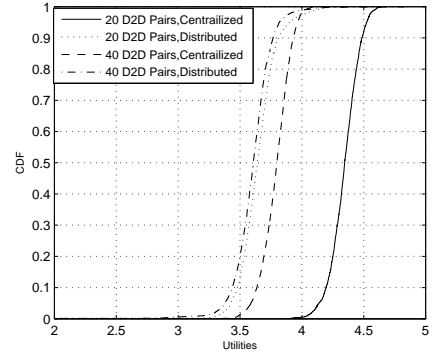}
\caption{The CDFs of Utilities of D2D pairs under different conditions with maximum distance between D2D TX and RX is 20m.}
\label{fig_sim1}
\end{figure}

Fig.\ref{fig_sim1} compares the utilities' cumulative distribution function (CDF) of D2D pairs among the centralized scheme solving by CVX and proposed distributed scheme. The result shows that there is about 10\% utility loss caused by the constraint of reusing the resource of only one cellular UE.
The gap between the centralized scheme and distributed scheme decrease when the number of D2D pair increases.

\begin{figure}[!t]
\centering
\includegraphics[width=2.5in]{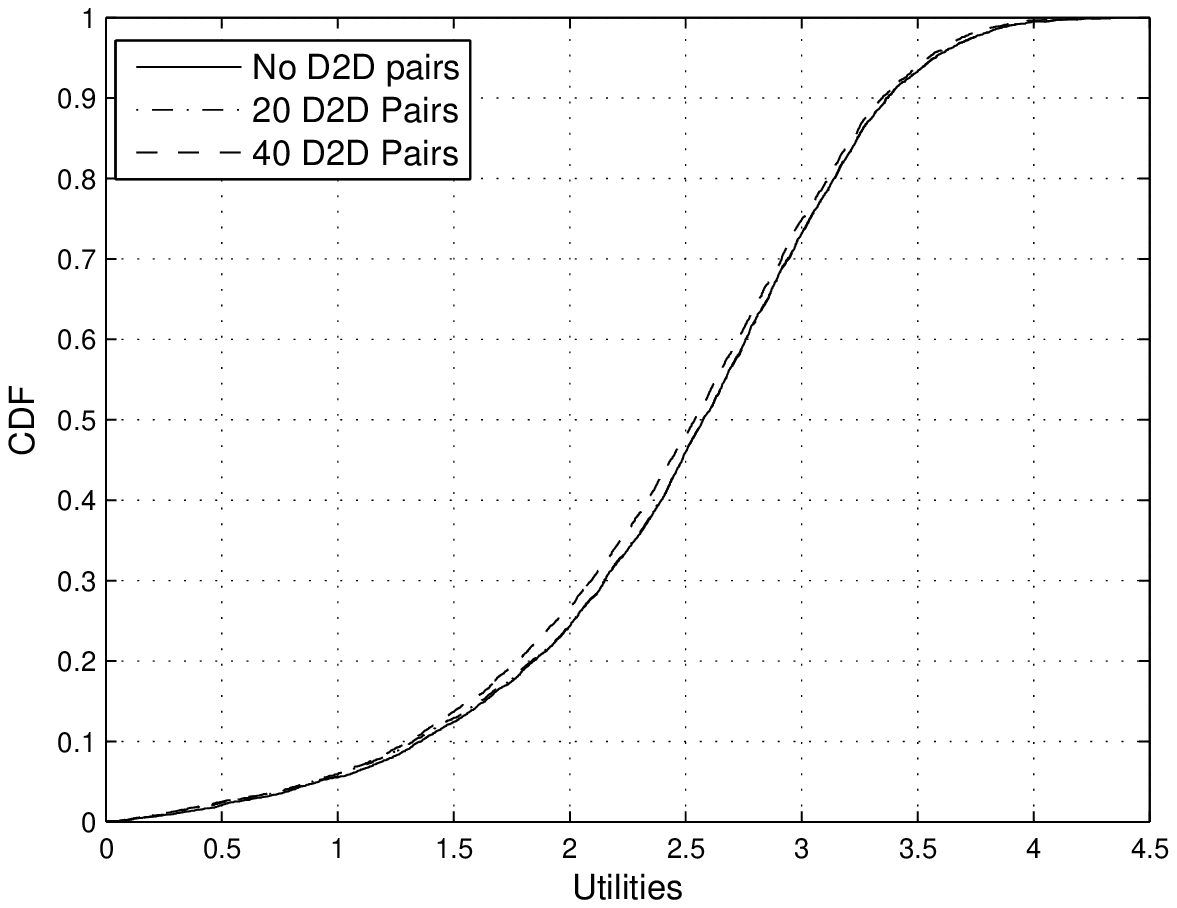}
\caption{The CDFs of Utilities of Cellular UEs under different conditions with maximium distance between D2D TX and RX is 20m.}
\label{fig_sim2}
\end{figure}

Fig.\ref{fig_sim2} shows the utilities' CDF of cellular UEs among different schemes.We can find that the three curves almost overlap which proves D2D communications underlaying cellular networks is a practical way to improve the system capacity.

\begin{figure}[!t]
\centering
\includegraphics[width=2.5in]{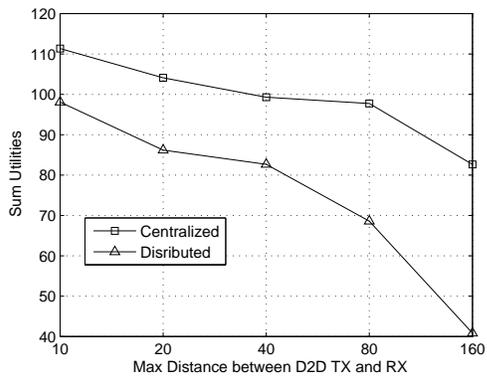}
\caption{Sum utilities of the system with different maximum D2D distances with 25 D2D pairs}
\label{fig_sim3}
\end{figure}

Fig.\ref{fig_sim3} shows the influences the maximum transmission distance between D2D Transmitter(Tx) and Receiver(RX).The maximum transmission is a key parameter of D2D communication underlaying cellular networks as there is less benefits from the lower pathloss between D2D TX and RX and more interference to the cellular network.The result show that the utilities goes down rapidly when the maxmium distance of D2D TX and RX increases.


\section{Conclusion}
In this paper, we studied D2D communications as an underlaying network to cellular communications. We formulated a binary combinational programming problem for resource allocations and relax it into a convex optimization problem aiming at the maximization of sum utilities of D2D users and satisfaction of the QoS requirement of cellular users. We further designed a distributed algorithm by using primal-dual decomposition. In this algorithm, we allow a D2D pair can share only one cellular user's resources to reduce the complexity. Finally the numerical result is given to demonstrate the effectiveness of the scheme.


\section*{Acknowledgement}
This paper is supported in part by National Basic Research (973) Program under Grant 2012CB316003, National High-Tech R\&D (863) Program under Grant 2012AA011402, and Doctoral Fund of MOE under Grant 20110185130003 of China.



\bibliographystyle{IEEEtran}
\bibliography{IEEEabrv,bare_conf}
%



\end{document}